\documentclass[11pt]{article}

\usepackage{setspace}
\usepackage{fullpage}
\usepackage{dsfont}
\usepackage[basic,physics]{circ}
\usepackage{authblk}
\usepackage{nomencl}

\usepackage{amssymb}
\usepackage{amsmath}
\usepackage{amsfonts}
\usepackage{graphicx,subfigure}
\usepackage{epstopdf}
\usepackage{color}
\usepackage{eso-pic}
\usepackage{etoolbox}
\usepackage{widetext}
\usepackage{caption}
\usepackage[hyphens]{url}

\newcommand{\be}{\begin{equation}}
\newcommand{\ee}{\end{equation}}

\begin{document}
\title{The Interhospital Transfer Network for Very Low Birth Weight Infants in the United States}
\author[a,b]{Munik Shrestha,\thanks{m.shrestha@northeastern.edu}}
\author[a,b,c,d]{Samuel V. Scarpino, \thanks{s.scarpino@northeastern.edu}}
\author[e,f,g]{Erika M. Edwards\thanks{Corresponding author: eedwards@vtoxford.org}}
\author[e]{Lucy T. Greenberg\thanks{lgreenberg@vtoxford.org}}
\author[e,f]{Jeffrey D. Horbar \thanks{ horbar@vtoxford.org} }

\affil[a]{Network Science Institute, Northeastern University, Boston, MA, 02115, USA}
\affil[b]{Marine \& Environmental Sciences, Northeastern University, Boston, MA, 02115, USA}
\affil[c]{Physics, Northeastern University, Boston, MA, 02115, USA}
\affil[d]{ISI Foundation, 10126 Turin, Italy}
\affil[e]{Vermont Oxford Network, Burlington, VT, 05401, USA}
\affil[f]{Pediatrics, University of Vermont, 05405, USA}
\affil[g]{Mathematics and Statistics, University of Vermont, Burlington, VT, 05405, USA}

\maketitle

\begin{abstract}
Very low birth weight (VLBW) infants require specialized care in neonatal intensive care units. In the United States (U.S.), such infants frequently are transferred between hospitals. Although these neonatal transfer networks are important, both economically and for infant morbidity and mortality, the national-level pattern of neonatal transfers is largely unknown. Using data from Vermont Oxford Network on 44,753 births, 2,122 hospitals, and 9,722 inter-hospital infant transfers from 2015, we performed the largest analysis to date on the inter-hospital transfer network for VLBW infants in the U.S.  We find that transfers are organized around regional communities, but that despite being largely within state boundaries, most communities often contain at least two hospitals in different states. To classify the structural variation in transfer pattern amongst these communities, we applied a spectral measure for regionalization and found an association between a community's degree of regionalization and their infant transfer rate, which was not utilized in detecting communities. We also demonstrate that the established measures of network centrality and hierarchy, e.g., the community-wide entropy in PageRank or betweenness centrality and number of distinct `layers' within a community, correlate weakly with our regionalization index and were not significantly associated with metrics on infant transfer rate.  Our results suggest that the regionalization index captures novel information about the structural properties of VLBW infant transfer networks, have the practical implication of characterizing neonatal care in the U.S., and may apply more broadly to the role of centralizing forces in organizing complex adaptive systems. 
\end{abstract}


\makenomenclature
\renewcommand{\nomname}{}
\setlength{\nomitemsep}{8pt}
\renewcommand\nomgroup[1]{%
  \item[\Large\bfseries
  \ifstrequal{#1}{N}{Nomenclature}{%
  \ifstrequal{#1}{A}{List of Abbreviations}{}}%
]\vspace{10pt}} 

\nomenclature[A]{\textbf{VLBW}}{Very low birth weight}
\nomenclature[A]{\textbf{US}}{United States}
\nomenclature[A]{\textbf{VON}}{Vermont Oxford Network}
\nomenclature[A]{\textbf{NICU}}{Neonatal intensive care unit}

\begin{spacing}{1.}
\section*{Introduction}

Although very low birth weight (VLBW) infants, i.e. individuals weighing less than 1,500 grams at birth, accounted for only 1.4\% of all births in the United States (U.S.) in 2015, they accounted for 52\% of all infant deaths~\cite{cdc2015wonder}. These extremely fragile infants require specialized care in a neonatal intensive care unit (NICU). 

The creation of regional systems for perinatal care, first proposed in 1975~\cite{ryan1975toward}, envisioned three levels of care including NICUs. The complexity of a patient's needs was meant to determine the hospital where the mother or infant received care. The goal was to improve outcomes for high-risk pregnant women and VLBW infants by ensuring access to high-quality, economically-efficient care for all mothers and their newborns~\cite{ryan1975toward}.

In the ensuing 40 years, however, systems of perinatal care developed based on financial incentives, geography, patient preferences, and hospitals' interests in establishing NICUs to attract young families~\cite{goodman2002relation, lorch2010regionalization}. This growth has led to the deregionalization of care, the proliferation of smaller maternity centers and NICUs, and the uneven distribution of perinatal resources unrelated to regional requirements~\cite{gould2002expansion,rayburn2012drive,kastenberg2015effect, harrison2017regional}. Regionalization models, regulations, and measures of risk-appropriate care for high risk infants vary widely among states~\cite{stark2004levels, lorch2010regionalization, nowakowski2012assessment, okoroh2016united, brantley2017perinatal,kroelinger2017comparison}.

One effect of deregionalization -- coupled with the unavoidable fact that emergency deliveries sometimes happen far from home -- is that VLBW infants are not always born at a hospital capable of providing the appropriate level of care for their entire stay, and require transport to a different facility. There is substantial evidence suggesting that birth at a hospital with an appropriate NICU and care at a NICU with an adequate volume of VLBW infants are associated with improved survival~\cite{mccormick1985regionalization,paneth1982newborn,rogowski2004indirect, stark2004levels, phibbs2007level,lasswell2010perinatal,lorch2012differential}.  

Recently, the American Academy of Pediatrics and the American College of Obstetrics and Gynecology developed new functional classifications of facilities that provide hospital care for pregnant women and newborn infants and recommend regionalized systems of perinatal care based on these classifications~\cite{stark2004levels, barfield2012levels,menard2015levels}. In the American Academy of Pediatrics guidelines, the highest level (III or IV) NICUs can care for VLBW infants throughout their entire birth hospitalization. Mothers should be transferred to such hospitals prior to birth, or VLBW infants born at a hospital with a lower level NICU should be transferred to a hospital capable of providing the appropriate level of care.  Understanding how hospitals work together in regionalized networks to care for infants is an important part of that effort~\cite{rashidian2014effectiveness,lorch2015ensuring}. In this study, we perform a network analysis on a U.S. national-level population of VLBW infants.  Although the application of network science methodologies to large, empirical patient transfer networks has only recently been possible, similar approaches have been previously applied to NICUs in California~\cite{kunz2017network} and adult intensive care units nationally~\cite{iwashyna2009structure}.

A rich literature on detecting the presence of hierarchical and/or centralization structure in networks exists, which could be extended to study regionalization. For example, methods exist to perform hierarchical decomposition and link prediction in complex networks~\cite{clauset2008hierarchical}, for identifying interactions between groups in multilayers networks~\cite{valles2016multilayer}, determining the flow-hierarchy of a network (defined as the fraction of edges in the community that are not in the cycle graph~\cite{jianxi-chris-2011}), and detecting overlapping communities in hierarchical networks~\cite{shen2009detect}. Additionally, fluctuations--or variation--in measures such as PageRank and betweenness centrality (both terms we define in the methods section) are often applied to measure hierarchy and centralization in networks \cite{Corominas-Murtra13316,Johnson5618,Corominas-Caso-2010}. Lastly, the role of hierarchical structure has also been studied in the context of gene interaction~\cite{yu2006genomic}, brain~\cite{meunier2010modular} and hydrological~\cite{lebecherel2016evaluating} networks. Our measure of regionalization closely maps on to established, network-science metrics for describing both the centralization and degree of hierarchical structure in networks.

Here, we apply a network science methodology to national data from Vermont Oxford Network. Our approach first identifies groups of highly connected hospitals, i.e. communities of hospitals, as measured by VLBW infant transfers. We then quantify the degree to which these communities are different in terms of regionalization by defining a regionalization index as the entropy in singular values of a normalized adjacency matrix. An important feature of our method is that it retains information about the frequency and direction of VLBW infant transfers, which advances our understanding of VLBW infant transfer networks and our capacity to study weighted, directed networks. By characterizing network regionalization, we find that the degree of regionalization within communities varies across the U.S. and that this variation is associated with key information on infant care.  Our results further highlight the growing utility of applying network science approaches to hospital care networks.

\section*{Results}
\begin{figure*}[ht]
\centering
\includegraphics[width=.71\linewidth]{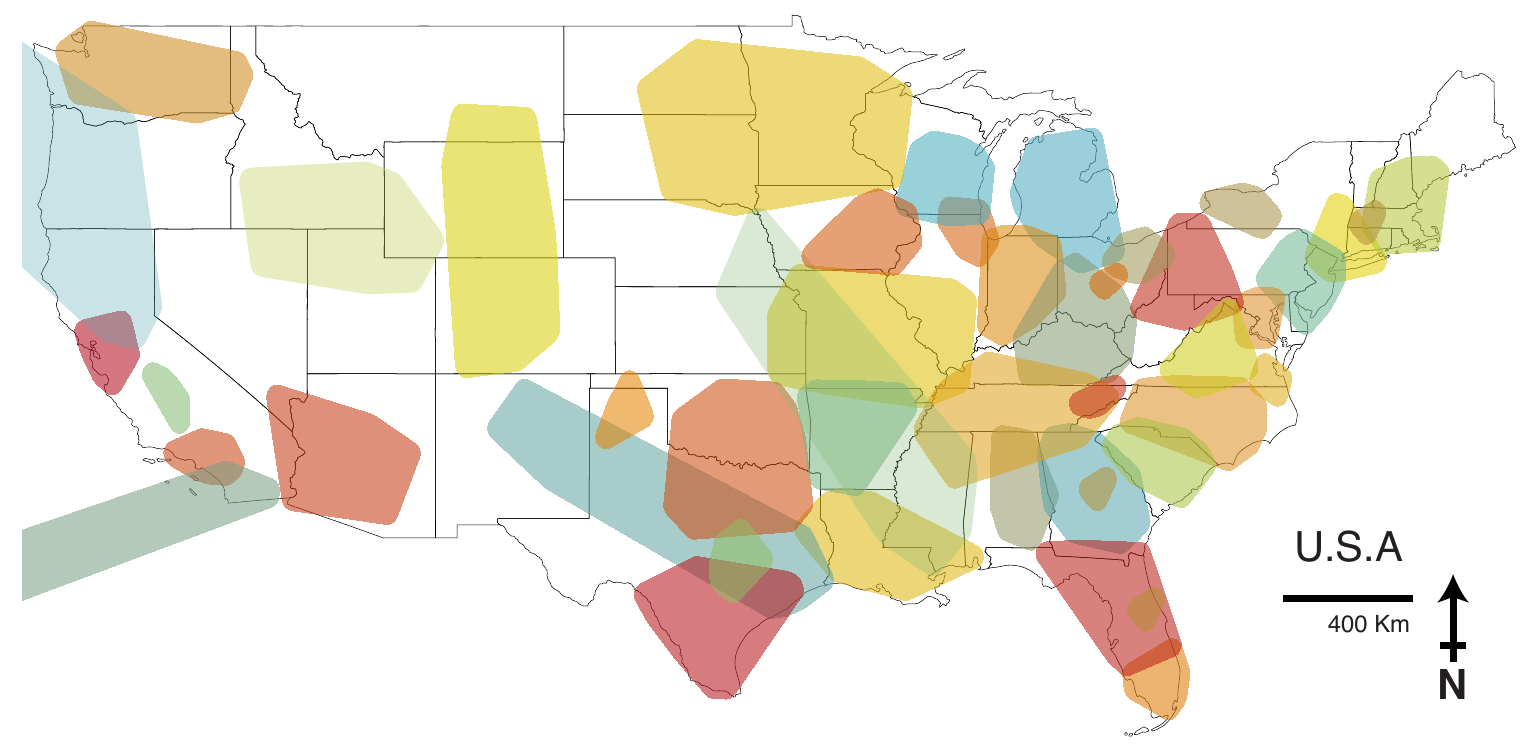}
\caption{The 2,122 hospitals were assigned to 50 statistically determined communities, i.e. a group of hospitals that transfers many more infants amongst each other than with hospitals outside of the ``community.''  Communities are represented by a minimum spanning polygon around 90\% of the infant transfers. The borders have been obscured to protect the privacy of the health care facilities.}
\label{fig-community}
\end{figure*}

Using data from 9,722 VLBW infant transfers among 2,122 hospitals in the U.S., we constructed a weighted, directed network. From this network, we identified groups, or communities, of hospitals that transferred VLBW infants more frequently among themselves than with other communities, see Figure~\ref{fig-community}. More specifically, we followed the established procedure of maximizing the number of within-community transfers relative to between-community transfers, i.e. modularity maximization. Given that the VLBW infant transfer network is directed and weighted by the number of infant transfers between hospitals, we applied Blondel and Guillaume et al.'s ~ approach to modularity maximization to search for high modularity partitions of the network \cite{Blondel2008}. 

Briefly, their approach is a bottom-up, multilevel clustering algorithm where each node is moved to other communities iteratively in order to maximize the local contribution to the modularity function. When the algorithm reaches a local maximum, all nodes within the community are mapped onto a single node--while tracking the weights of associated edges--and the process of searching for better modularity scores continues to the next ``level." As a result, the algorithm naturally identifies both the communities and the number of hierarchical layers in each community.  However, as we outline below, the number of ``layers'' detected via this method provided an overly coarse description of the VLBW infant transfer networks, which was largely redundant with established centrality measures, e.g., PageRank and betweenness centralities.  

\begin{table}[ht]
\caption{Hospital and community characteristics with the interquartile range (IQR), i.e. the interval between the 25th and 75th percentiles.}
\centering
\begin{tabular}{l c c c}
\hline \hline
  & Overall & By community & By hospital\\
  &  & \textit{Median (IQR)} &\textit{Median (IQR)}\\[0.7ex]
 \hline
Infants (total) & 44,753 & 787 (400-1,325) & 51 (25-95)\\
Infants transfered & 9,722 & 178 (80-272) & 11 (6-19)\\
Hospitals & 2,122 & 40 (22-63) &  --- \\
Level III or IV hospitals & 27\% & 25\% (17-33\%) & --- \\
States covered (inc D.C.) & 51 & 2 (1-4) & --- \\ [1ex]
\hline
\end{tabular}
\label{table:infants}
\end{table}

Using this community detection algorithm, we identified 53 communities within the U.S. interhospital transfer network for neonates. To protect patient privacy, we excluded networks with $<3$ infant transfers and we manually split a community, which included hospitals in Georgia and Connecticut. As a result, we used 50 distinct communities in our analysis.  Figure \ref{fig-community} shows the geographic distribution of these communities. Out of 9,722 transfers, there were only 222 transfers (or 2.3\%) between hospitals in different communities. The communities varied in the number of hospitals, number of infants, and percentage of infants in the community who were transferred, see Table~\ref{table:infants}. Thirty-two of 50 communities had hospitals in more than one state; nevertheless, less than 8\% of transfers occurred between hospitals in different states. This feature--where few transfers occur between states, but most communities contain at least one hospital in different states--arises because the average number of transfers from an individual hospital is quite low (median 11 (6-19 IQR), see Table~\ref{table:infants}) relative to the $>700$ (or approx. 8\%) of inter-state transfers. Nevertheless, developing a fuller understanding of the mechanisms behind our observation that inter-state transfers are rare remains an important area of future research. 

\subsection*{Structural variation within VLBW communities}
To measure structural variation across these 50 communities, we employ a continuous index for the degree of network regionalization. More formally, for a graph $G$, defined by a normalized adjacency matrix $A$ whose elements $A_{ij}$ are the fraction of transfers from hospital $j$ to hospital $i$, we define regionalization $\chi (G) \rightarrow \mathbb{R^+}$ as the  entropy in the normalized singular values of $A$.  Note then that the entropy is minimum for star-graph, i.e., $\chi(G)=0$, and maximum or $\chi(G)=\log(N)$ for cycle graphs separating star-like graphs from cycle graphs. We provide more details on the computation of this metric--along with its performance on stylized networks--in the methods section.

\begin{table}[ht]
\centering
\caption{Table}{Network Metrics by Community. (\textit{Network measures are defined in the methods.})} 
\resizebox{\columnwidth}{!}{%
\begin{tabular}{l c c c c c c}
\hline\hline
 & Regionalization $\chi(G)$ & Hierarchical layers & PageRank & Betweenness & Reciprocity ($\alpha$) & Flow Hierarchy ($\eta$) \\ [0.7ex] 
\hline
Median & 2.22 & 2 &    2.77 &    1.34 &    0.19 &    0.77\\
Mean &   1.99 &    2.28 &   2.62 &     1.29 &     0.20 &     0.75  \\
I.Q.R &  (1.48 - 2.63) & (1 - 3) & (2.05 - 3.27) & (0.86 - 1.74) & (0.11 - 0.25) & (0.60 - 0.88) \\
Min.\slash Max. & (0 - 3.50) & (1 - 6) & (0.65 - 3.77) & (0.00 - 2.66) & (0.00 - 0.63) & (0.31 - 1.00) \\[1ex]
\hline
\end{tabular}%
}
\label{table:network}
\end{table}

From a network science perspective, the regionalization index provides both a measure of how fast a random walker moving through the graph mixes to its stationary distribution and how fast groups of nodes (as indicated from the component of singular vectors) mix to the leading singular value. This approach of including the whole spectrum of the graph resonates with analogous spectral methods used to detect synchronization time scales in hierarchical networks \cite{PhysRevLett.96.114102}. However, VLBW infant transfer networks are weighted, directed, asymmetric, and have adjacency matrices with complex eigenvalues. To simplify the analysis, while still preserving aspects of the relevant structure, we applied a straightforward transformation, $A^TA$, to construct a symmetric matrix where all of the eigenvalues are real. These eigenvalues are also the square of the singular values of $A$ and thus preserve the relevant structure of the original graph.  

\begin{figure}[ht]
\centering
\includegraphics[width=.61\linewidth]{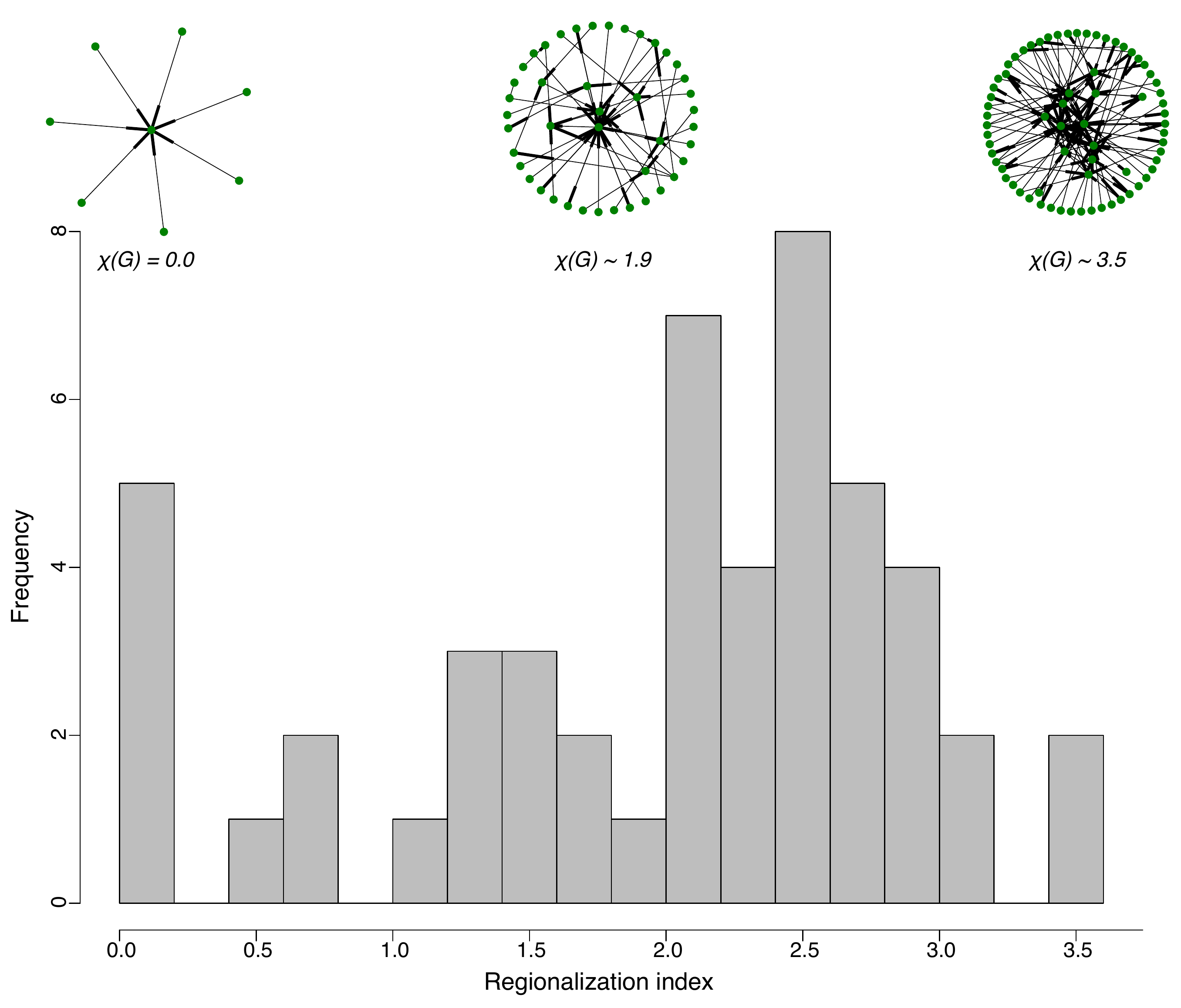}
\caption{The distribution of regionalization indices across the 50 U.S. VLBW infant transfer network communities.  A score of 0 is associated with the most centralized networks, as represented by the empirical network in the top-left of the plot, while a score of 3.5 represents the most regionalized empirical network identified in our study, as represented by the network in the top-right.  The mean observed regionalization index observed across the VLBW infant transfer network communities was 1.99, which closely matches the empirical network at $h = 1.9$.}
\label{fig-regionalization}
\end{figure}

After calculating the regionalization index, $\chi(G)$, across all 50 VLBW infant transfer communities, we find that on average networks have a regionalization index of 1.99, ranging from exactly 0 to 3.5, see Figure~\ref{fig-regionalization}. Interestingly, five communities had regionalization scores of exactly 0, meaning that they are perfect star graphs, or hub-and-spoke networks. In these communities, there was a single level III or IV hospital that received infants from multiple, lower-level hospitals. 

\begin{figure}[ht]
\centering
\includegraphics[width=.61\linewidth]{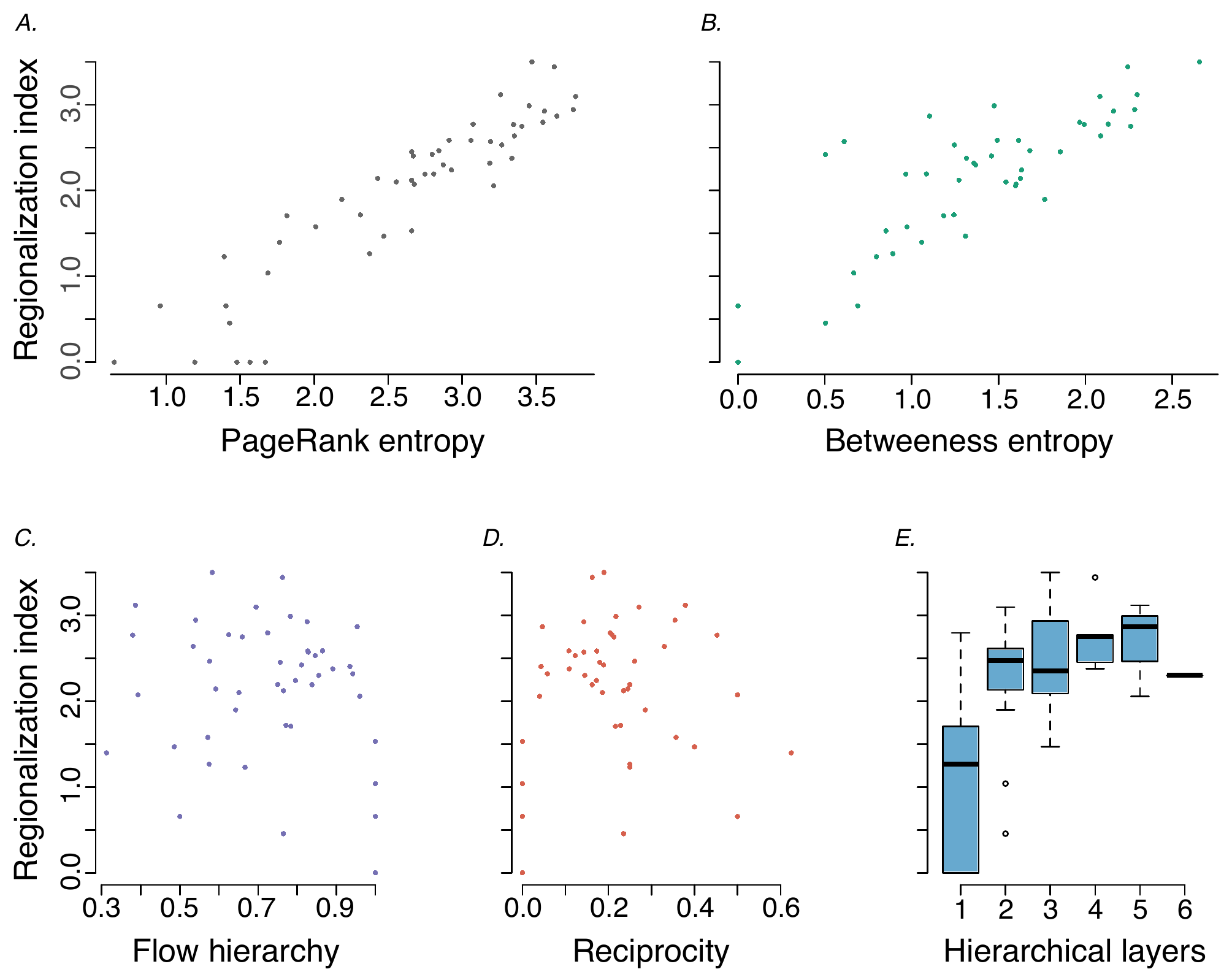}
\caption{The association between the regionalization index and the community-level \textit{A.} PageRank entropy, \textit{B.} betweenness centrality entropy, \textit{C.} flow hierarchy, \textit{D.} reciprocity, and \textit{E.} number of hierarchical layers. The regionalization index was positively associated with the community-level entropy of PageRank ($R^2 = 0.85$; $p < 0.001$) and betweenness centrality ($R^2 = 0.76$; $p < 0.001$); however, regionaliZation was negatively associated with flow hierarchy ($R^2 = 0.10$; $p < 0.014$), not linearly correlated with reciprocity ($R^2 = 0.05$; $p = 0.07$), and only weakly associated with the number of hierarchical layers (the mean regionalization index was significantly lower in communities with only a single hierarchical layer; ANOVA: degrees of freedom = 5;  F value = 7.245;  $p < 0.001$; Tukey HSD test, with a correction for multiple comparisons, was used post-doc to determine which group means were different).}
\label{fig-metricComparison}
\end{figure}

The VLBW transfer communities also varied across established network science hierarchy metrics \cite{Corominas-Murtra13316,Johnson5618,Corominas-Caso-2010}, as measured by the community-level entropy in PageRank and betweenness Centrality, the flow hierarchy, and the number of hierarchical layers, Table~\ref{table:network} and Figure~\ref{fig-metricComparison}. Using the proportion of variation explained in an ordinary least-squares regression, $R^2$ as our measure of association, the regionalization index was positively associated with the community-level entropy of PageRank ($R^2 = 0.85$; $p < 0.001$) and betweenness centrality ($R^2 = 0.76$; $p < 0.001$); however, regionalization was negatively associated with flow hierarchy ($R^2 = 0.10$; $p < 0.014$), not linearly correlated with reciprocity ($R^2 = 0.05$; $p = 0.07$), and only weakly associated with the number of hierarchical layers (the mean regionalization index was significantly lower in communities with only a single hierarchical layer; ANOVA: degrees of freedom = 5; F value = 7.245; $p < 0.001$; Tukey HSD test--with a correction for multiple comparisons--was used post-hoc to determine which group means were different). However, as can be seen in Figure~\ref{fig-metricComparison}, the relationship between these metrics -- especially reciprocity and the regionalization index -- cannot be fully described with a linear model. Lastly, the average PageRank centrality (along with other measures of centrality) was higher among level IV NICUs, as compared to lower-level facilities (ANOVA: degrees of freedom = 3; F=106.1; $p<10^-5$; with a TukeyHSD post-hoc test to determine that level 4 NICUs were driving the difference in means). However, we leave a richer exploration of differences between the level of care a hospital can provide and transfer network metrics for future research.

\begin{table}[ht]
\centering
\captionof{table}{Association between \textit{proportion not transfered} and community-level network metrics. In the regression models, we controlled for variation in the number of hospitals and the fraction of VON member hospitals within each community.}
\centering
\begin{tabular}{lccc}
  \hline \hline
 Metric & Adjusted $R^2$ & $p$ value\\ 
  \hline
	Regionalization Index & 0.346 & $<0.001$ \\ 
    PageRank Entropy & 0.158 & 0.085 \\ 
    Betweenness Entropy & 0.164 & 0.071 \\ 
    Flow hierarchy & 0.129 & 0.228 \\
    Reciprocity & 0.152 & 0.103  \\[1ex]
   \hline
\end{tabular}
\label{table:association}
\end{table}

\begin{figure}[ht]
\centering
\includegraphics[width=13cm]{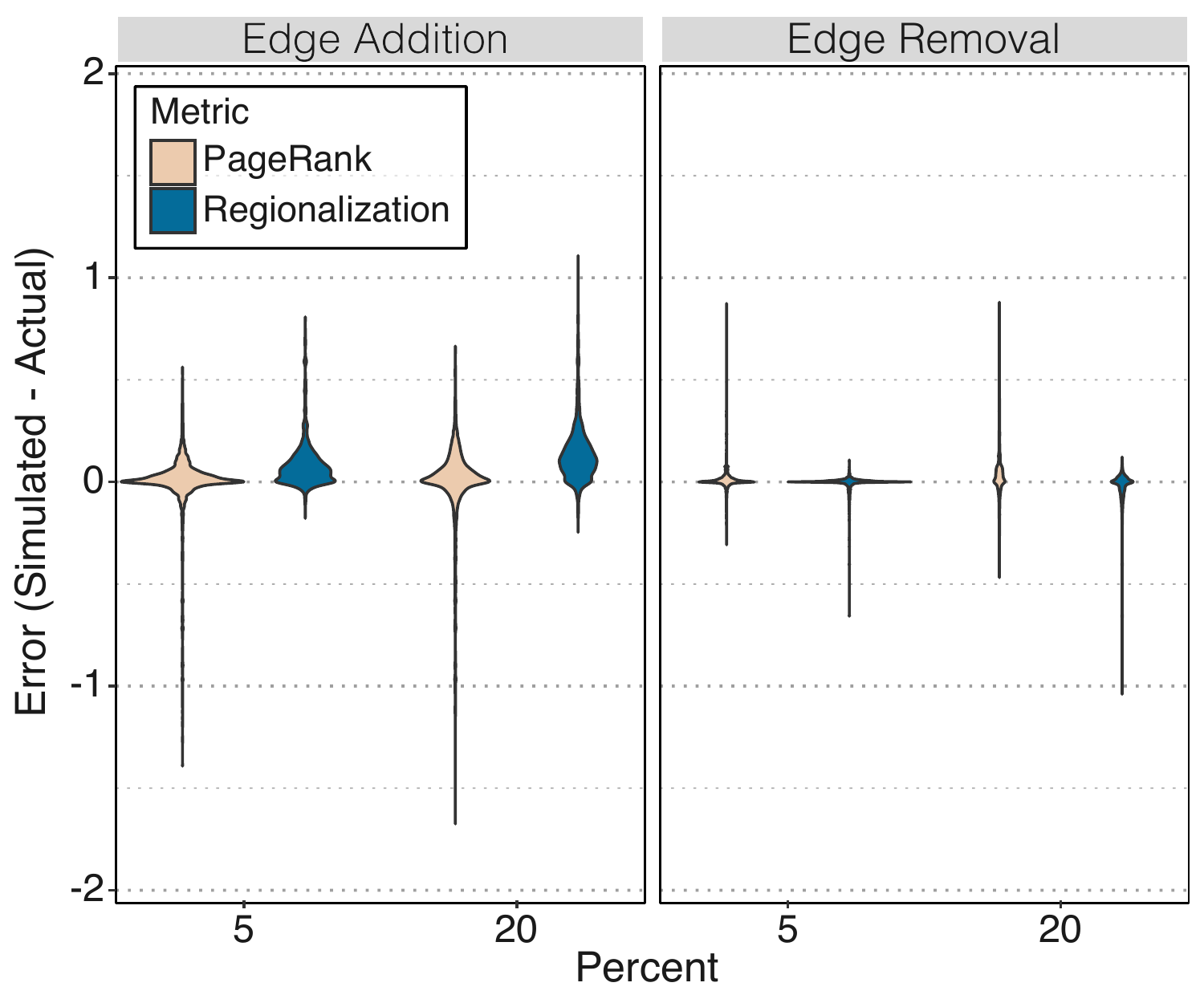}
\caption{\small Average (over 100 simulated networks for both edge addition and edge removal) PageRank entropy (tan) and regionalization index (blue) when either 5 or 20 percent of the edges were randomly added (left panel) or removed (right panel) in proportion to edge weights in the original network. The bias for both metrics was nearly 0, with PageRank performing slightly better during random addition and regionalization performing slightly better during random deletion. These differences were statistically significant by ANOVA with a post-hoc Tukey Honest Significant Differences test and false discovery correction to control for multiple comparisons.} 
\label{fig-region-pagerank}
\end{figure}

Although our data set contains nearly 90 percent of the VLBW infants born in the US, i.e. it was approximately 90\% complete, we assessed the robustness of our methodology to missing data by performing edge addition\slash deletion.  Briefly, we randomly increased (or decreased) the number of transfers between hospitals by either 1\%, 5\%, or 20\%, proportional to the prevalence of transfers in the original network and, in addition, added a small amount of uniform noise to the edge weights.  With these ``simulated'' networks, we both re-estimated the community structure (although not for 20\%) and, using the original community structure, estimated the entropy over the PageRanks and the regionalization index for each community. For community detection, we found that--even with 5\% addition\slash removal of transfers--more than 90\% of hospitals were grouped in the same community as in the original network and that the error for both the PageRank entropies and regionalization indexes was nearly zero (in this case even up to 20\% addition\slash removal), see Figure~\ref{fig-region-pagerank}.  Interestingly, the PageRank entropy was slightly more robust to edge addition, while the regionalization index was slightly more robust to edge deletion (differences were statistically significant by ANOVA with a post-hoc Tukey Honest Significant Differences test and false discovery correction to control for multiple comparisons).  Our results--general robustness of community detection and regionalization metrics to edge addition\slash removal, but a slight trade-off in the robustness of our two primary measures of regionalization--highlight the importance of considering multiple metrics when analyzing a data set and the utility of the Blondel and Guillaume et al. procedure for community detection in hierarchical networks.

\begin{figure}[ht]
\centering
\includegraphics[width=.61\linewidth]{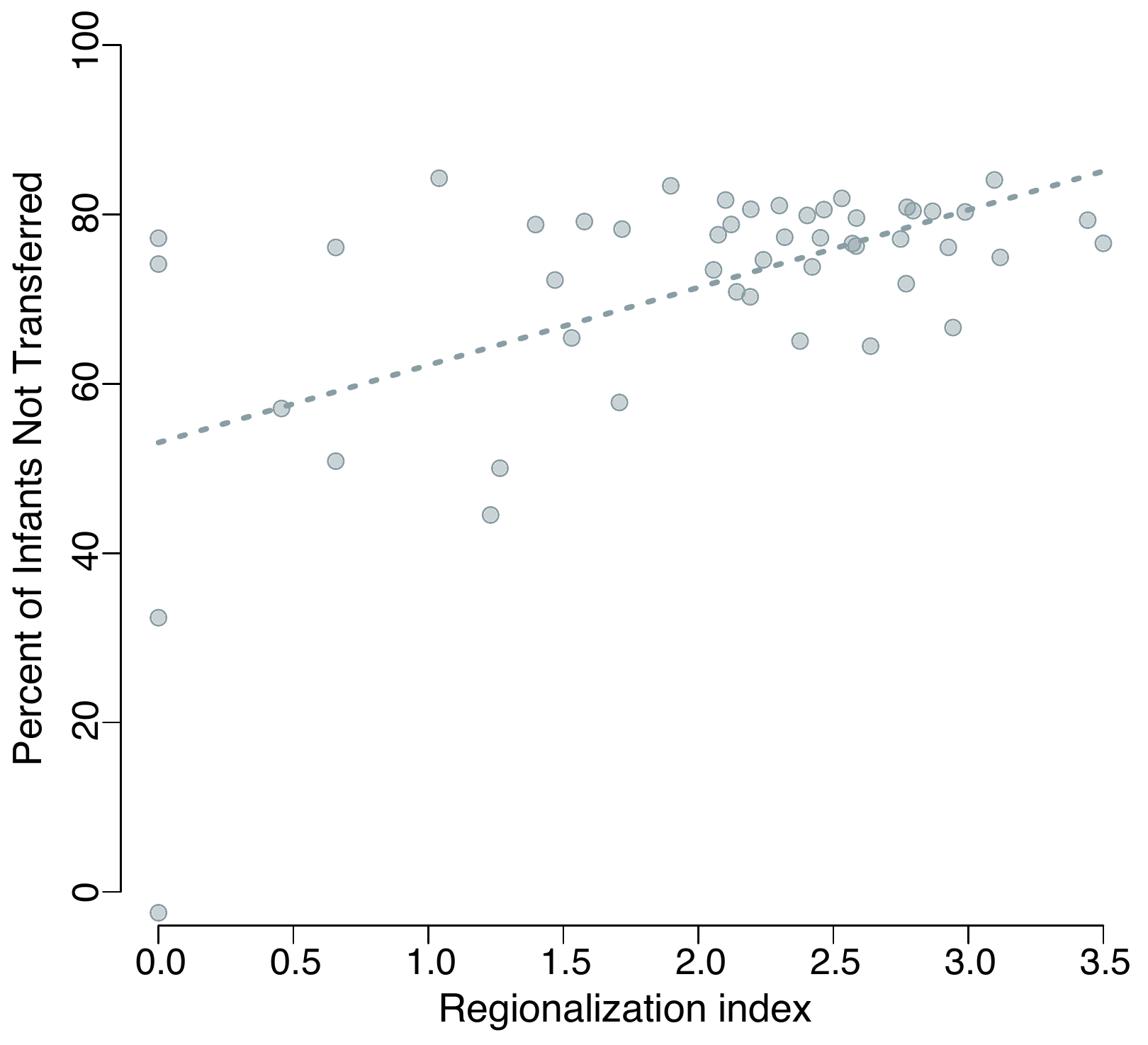}
\caption{The regionalization index and the \textit{proportion not transfered} are positively associated. We performed an ordinary least-squares regression between the regionalization index and the proportion of all infants born in a given community who were never transferred.  In this regression, we controlled for the total number of hospitals in each community and the fraction of each community that were VON member facilities (we only had data on the \textit{proportion not transfered} for VON member hospitals). The results were robust to removal of the point in the lower-left corner where there were all infants were transfered and the regionalization index was zero (regression results with ``outlier" removed: adjusted $R^2 = 336$; $p < 0.001$).} 
\label{fig-notTransfered}
\end{figure}

Lastly, we evaluated the relationship between the regionalization index and the proportion of all infants born in a given community who were never transferred (\textit{proportion not transfered}). These non-transferring infants neither contributed information directly to the community detection nor to the regionalization index. However, we found a positive association between the overall percent of infants who did not transfer and the hierarchical index for the community (adjusted $R^2 = 0.346$; $p < 0.001$), Figure~\ref{fig-notTransfered}. This result was robust to variation in the number of hospitals belonging to each community and to the fraction of VON hospitals present in the community (we only had data on the \textit{proportion not transfered} for VON member hospitals). Importantly, although the association was also positive, we failed to find statistically significant relationships between the \textit{proportion not transfered} and the other network methods, i.e. PageRank entropy, Betweeness entropy, flow hierarchy, reciprocity, and the number of hierarchical layers (ANOVA: degrees of freedom = 5;  F value = 1.795;  p = 0.135), see Table~\ref{table:association}.  This result, that the regionalization index has a significant relationship with the \textit{proportion not transfered}, further highlights the additional information about the network being captured by the regionalization index.

\section*{Discussion}
Using a U.S. national-level data set containing 9,722 VLBW infant transfers among 2,122 hospitals, we found a variety of different types of communities, varying from highly centralized to highly regionalized. That we found strongly regionalized networks is perhaps not surprising given the financial incentives of the health care market, the differences in state regulations governing NICU expansion, and geography~\cite{goodman2002relation,lorch2010regionalization,nowakowski2012assessment,okoroh2016united,brantley2017perinatal,kroelinger2017comparison,rayburn2012drive}. Our communities were largely organized around state boundaries, but 25\% of communities overlapped at least four states and 8\% of all transfers occurred between hospitals in different states. The interstate nature of transports highlights the importance of states working together to coordinate policies in these regional NICU networks to improve efficiency and quality of care. 

To obtain these results on VLBW infant transfer network structure, we developed and applied a network science measure able to distinguish networks based on their degree of regionalization, which also accounts for the weighted, directed, and asymmetric nature of VLBW transfers. We verified this metric using data from all infants in the detected communities, finding the percent of infants not transferred was associated with greater regionalization scores. 

Our study differs from Kunz et al.~\cite{kunz2017network}, who characterized NICU referral networks in California, and Iwashyna et al.~\cite{iwashyna2009structure}, who described adult critical care transfer networks nationally. Kunz et al. collapsed all transfers between hospitals into a single edge and made the network symmetric by removing the directionality of transfers, then used the Fortunato community detection algorithm~\cite{fortunato2010} to identify communities, while Iwashyna et al. used PageRank~\cite{ilprints422} to calculate centrality. These methods, while valid, do not make use of all of the information provided by a hierarchical network in which transfers are weighted, directed, and asymmetric. Additionally, given our findings that many VLBW infant transfer communities span multiple states, obtaining an accurate picture of network structure necessitates national-level data.  Therefore, our results likely provide the first picture of national-level VLBW infant hospital transfer networks. 

There is a rich and growing body of literature on detecting hierarchical structure in networks~\cite{krackhardt2014graph, trusina2004hierarchy, mones2012hierarchy}.  Our work advances from earlier studies on identifying the presence of hierarchy in networks by defining a metric of regionalization structure, which provides a fuller description of the network than many of the standard centrality measures.  Briefly, Corominas-Murtra (2013) maps the definition of hierarchy to three dimensional points (Orderability \textit{O}, Feedforwardness \textit{F} and Treeness \textit{T}), Luo \& Magee (2011) defines hierarchy as the fraction of edges that are not a part of the cycle graph, and Cz\'egel \& Palla (2015) measure hierarchy as the normalized variance in the stationary distribution of a random walker on a network~\cite{Corominas-Murtra13316, jianxi-chris-2011, czegel2015random}. Clearly, Cz\'egel and Palla's approach is closely related to ours and will be nearly identical to the PageRank entropies we report for each community~\cite{czegel2015random}. Most similar to our work is a methodology proposed for using eigenvalues to reveal the time-scale of synchronizing dynamics in hierarchical networks \cite{PhysRevLett.96.114102}. Our approach is able to detect regionalization structure for asymmetric, weighted, and directed graphs. Additionally, the regionalization index we derived provides novel structural information about a network, as compared to established network science metrics.  Critically, our regionalization metric captured a relationship between network structure and the fraction of infants who are born into a hospital community but never transfered, which was not captured by the established metrics used in our analysis, which suggests that the regionalization metric contains novel information about network structure.

In 2015, VON members submitted data on nearly 90 percent of the VLBW infants born in the US. Most VON members are Level III or IV hospitals; however, in this analysis, members tracked the exact names of hospitals where infants originated or were sent, which included non-VON members. Although non-VON members do not report data to VON, we did know their levels of care. The only NICUs completely absent from our analysis were non-VON members that never transferred infants or exclusively transferred infants to other non-VON members. Since the VON data contains information on nearly all VLBW infants, an interesting avenue of future research would be to compare the nearly complete VON data to alternative, but less complete, data sets. Medicaid would provide information on the entire U.S., and as we have shown, understanding how communities cross state lines is vital. Additionally, many areas have systems of care where infants are born at a hospital capable of providing the appropriate level of care, or where transfers are not an option. We need to understand the systems of care in areas that do not have transfer communities and compare infant outcomes between different types of systems. Insurance markets inevitably play a large role, another area for future research.

One aspect of the VLBW transfer network that could interest the broad network science community -- and separates these networks from physical structures such as the power grid or hydrological networks -- is that transfers largely exist because infants were transferred up to a higher level of care, transfered laterally to a hospital with a similar level of care but different services, or were transferred down to a lower level of care before discharge home. As a result, from a complex systems perspective, the VLBW infant transfer network is best considered an emergent property of bottom-up (e.g., hospital capacity) and top-down (e.g., state-level policies on  transport reimbursement) processes, but one that may still be strongly constrained by geography. Therefore, an implication of our results is that both top-down and bottom-up effects might have measurable impacts on emergent network structure, a debate currently ongoing in both the ecological~\cite{lynam2017interaction} and complex adaptive systems~\cite{flack2017coarse} literature.

To advance our understanding of top-down vs. bottom-up processes in hospital transfer networks more specifically, and complex systems more generally, an important extension of this work is to analyze how the VLBW infant transfer network structure has changed over time; in particular, whether changes in state-level policies on transport or changes in reimbursement and financial incentives are associated with the types of changes in network structure predicted by our results. However, one important policy implication of the current research shows that VLBW transfers cross state borders while states policies\slash regulations governing maternal and neonatal transport do not~\cite{okoroh2016united}. The associations between existing state laws and community structure remain an important part of future research.   

\section*{Conclusions}
This study represents the first attempt to analyze the weighted, directed, and asymmetrical nature of VLBW transfers in the U.S. We developed and applied a spectral hierarchy measure in networks, which we termed a regionalization index, and found that regionalization correlated with empirically known information about infants in the detected communities. While there is still more to learn about perinatal care networks, our results contribute to what is known about the organization of neonatal care in the U.S. and may more broadly apply to the role of hierarchical forces in organizing complex adaptive systems.  

\subsection*{Data source: Vermont Oxford Network}
Vermont Oxford Network is a voluntary collaboration of health care professionals around the world dedicated to improving the quality, safety, and value of care for newborn infants and their families~\cite{Horbar201029}. More than 700 NICUs in the U.S. participate in the Vermont Oxford Network database. Vermont Oxford Network members submit standardized data for infants with a birth weight of 401 to 1500 grams or a gestational age of 22 weeks 0 days to 29 weeks 6 days who are born in the member hospital or admitted within 28 days of birth without first having been discharged home. 

For this study, we included infants who were born from January 1, 2015, to December 31, 2015, submitted by 702 hospitals in the United States.  The additional 1,420 hospitals included in our analysis were not members of VON, but transfered an infant to and\slash or received an infant from VON member hospitals.  All hospitals that contributed finalized data for the study period were included. Local staff collected data using uniform definitions that did not change during the study period ~\cite{Horbar201030}. All data underwent automated checks for quality and completeness at the time of submission. The University of Vermont and Northeastern University Committees on Human Research considered use of the Vermont Oxford Network Research Repository not human subjects research for this study. Consent for the Research Repository was waived by the University of Vermont Committee on Human Research and the Northeastern University Institutional Review Board. All data were de$-$identified. 

Members submitted the exact name, city, and state of sending or receiving hospitals for infants who were transferred into or out of their hospitals. The number of transfers in this report includes all transfers to a VON member hospital within 28 days of birth, or from a VON member hospital on or before first birthday and prior to being discharged home. Infants transferred between two VON members within 28 days are counted at the receiving hospital. Multiple eligible transfers of a single infant are counted separately. Transfers are not included if the sending or receiving hospital is not recorded. For a complete list of participating facilities, please see Appendix A.

\subsection*{Network methods}
Here we provide additional details on the network methods used in our analysis.  

\paragraph*{Community detection and modularity}
The network is first partitioned into groups, i.e. communities, such that there were relatively more edges connecting the hospitals within the same group than there were with other groups. We apply a well practiced method for partitioning the networks into groups, which is known as modularity-maximization~\cite{Girvan11062002}. Mathematically, modularity $M (\mathbf{s}):   \mathbb{Z}_q^n   \rightarrow \mathbb{R}$ is a fitness function 

\be\label{modularity}
M (\mathbf s)= \frac{1}{m} \sum_{i,j} \left(a_{ij} -p_{ij} \right ) \delta(s_i, s_j)
\ee
of the partition ${ \mathbf{s} \in \mathbb{Z}_q^n }$ of all $n$ nodes into $q$ groups, i.e., partition where each node belongs to group $\{s_i \in {1,...,q}\}$, where $m$ is the number edges, $a_{ij}$ is the number of transfers from $j$ to $i$, and $\delta(s_i, s_j) $ is 1 if $s_i=s_j$ or 0 otherwise. Here,  $p_{ij}$ is the probability under the null distribution that there is an edge between $i$ and $j$. The intuition behind modularity maximization is that we are comparing the fraction of edges within the same group from the resulting partition of the network to the average fraction of edges within same group in an ensemble of randomly generated networks, i.e., when the null model distribution results from the values $p_{ij}$. 

An alternative interpretation of the values $p_{ij}$ is that they represent the parameters associated with the well-known configuration model.  The configuration model is a procedure for generating random instances of a network by swapping edges and holding the degree (number of neighbors) constant. In such randomly generated networks, two nodes $i$ and $j$ will be connected with probability proportional to their degree, e.g., if these nodes have degree $d_i$ and $d_j$ respectively, they will be connected with probability $$p_{ij}=d_i d_j /2m.$$ However, note that we do not have to perform this randomization to compute $p_{ij}$ since we can directly calculate $d_i$ and $m$.\\

\paragraph*{Hierarchical layers}
We applied Blondel and Guillaume et al.'s approach to modularity maximization to find communities~\cite{Blondel2008}. One immediate application of their approach is that it naturally outputs the number of ``layers" present in each community. In other words, for each community, to determine the number of ``layers," we simply count the number of times the Guillaume et al. algorithm identified a local maxima and thus collapsed all nodes into a single hierarchical layer.  Critically, this approach does not require pre-specifying the number of communities, nor does it require optimizing a set of tuning parameters.

\begin{figure}[ht]
\centering
\includegraphics[width=.67\linewidth]{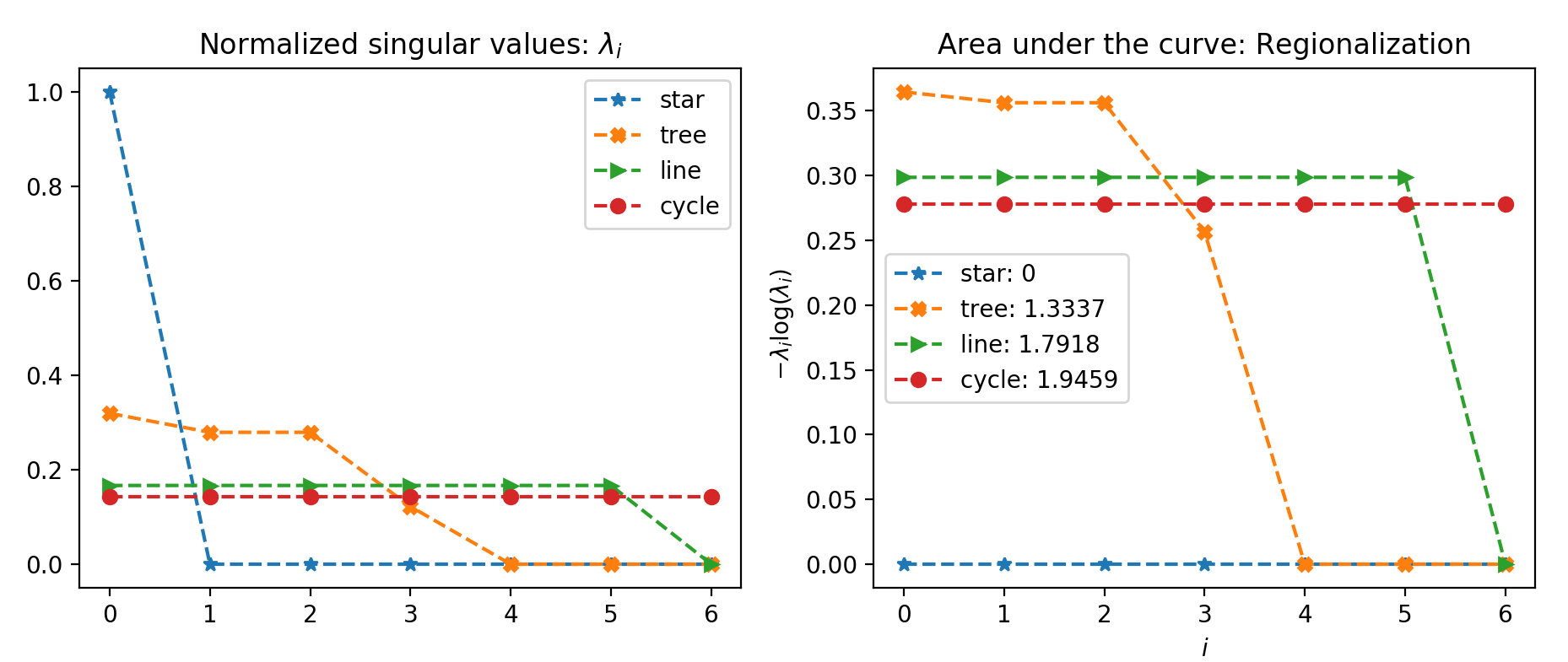}
\caption{An example computation of the regionalization index ($\chi(G)$) across a set simple, stylized graphs.  On the left, we show the normalized list plot of singular values of a star (blue-star), tree (orange-cross), line (green-triangle) and cycle (red-circle) graph with seven nodes each. On the right, we show the entropy list plot for each graph, i.e., the normalized singular values, and $\chi(G)$ is exactly the area under this curve (corresponding to values as listed in the caption).}
\label{fig-singulars}
\end{figure}

\paragraph{Regionalization Index}
We define regionalization $\chi(G)$ in a network as the entropy 
\be
\chi(G)= \sum_i \lambda_i \log \lambda_i,
\ee
in the normalized singular values $\lambda$ of an adjacency matrix. What we mean by normalized singular values is that sum of all singular values is 1.  Note that this entropy is minimum for star-graphs,i.e., $\chi(G)=0$ , and is maximized or $\chi(G)=\log(N)$ for cycle graphs, see Figure~\ref{fig-singulars} for an example calculation. 

\paragraph*{PageRank Centrality}
PageRank centrality was originally developed by Page et al \cite{ilprints422}  
as scalable algorithm to rank web pages based on the network structure of links to and from the web pages. Its centrality score for node $i$ is proportional to how often a random walker is at $i$, given it could either randomly, with probability $\gamma$, go to nodes that it points to, or with probability (1-$\gamma$) it could hop to any node in the graph with equal probability. In our calculation, $\gamma =0.85.$   The entropy in PageRank then relates to the uncertainty of a random walker undertaking such a walk in the network.  In our study, we calculate the PageRank entropy within a community, so random walkers would be restricted to moves, which keep them in the same community. 

\paragraph*{Betweenness Centrality}
Whereas PageRank centrality is based on a random walker following random paths through the network, betweenness centrality is computed from deterministic walks between all pairs of nodes.  More specifically, the betweenness centrality $b_i$ of node $i$ is the fraction of shortest paths between all pair of nodes in the network that pass through node $i$. In other words, the betweenness of node $i$, i.e. $b_i$, is the probability that the shortest path between a randomly selected origin node $j$ and destination node $k$ will ``pass through" node $i$. Again, the entropy in betweenness then relates to uncertainty in a random walker undertaking such a walk in the network. Similar to the PageRank entropy, we calculate the betweenness entropy within a community.

\paragraph*{Flow hierarchy} Flow hierarchy $\zeta$ of a network, is computed exactly as defined in \cite{jianxi-chris-2011}, but very briefly is the fraction of edges, which are not in a cycle. 

\paragraph*{Reciprocity} Reciprocity is the fraction of edges that are of cycle of length two. Note that reciprocity is related to the flow hierarchy because the fraction, $\theta$, of edges that are in cycles of length greater than two can be computed as $\theta=1-\zeta-\alpha$.

\end{spacing}
\printnomenclature[2cm]

\section*{Availability of data and material}
The infant-level data, including data on transfers, is protected by U.S. privacy laws and legal membership agreement. Individuals interested in using VON data are asked to follow our Policy and Guidelines for Collaborative Research using the Vermont Oxford Network Databases, available at \url{https://public.vtoxford.org/wp-content/uploads/2017/02/Vermont-Oxford-Network-Policy-and-Guidelines-for-Collaborative-Research-2017.pdf}. To further strengthen both the utility and reproducibility of our study, we have posted code associated with our analyses to the following github repository: \url{https://github.com/Emergent-Epidemics/VON_NICU_2018}.

\section*{Contributions}
All authors collaboratively designed and performed the research, contributed new analytic tools, analyzed data, and wrote the paper. 

\section*{Conflicts}
The authors declare no conflicts of interest exist.

\section*{Acknowledgements}
We thank Vermont Oxford Network members that contributed data to this study. We acknowledge our colleagues in the Network Science in Neonatology Working Group for their valuable discussions that deepened our understanding of the applications of network science in neonatology: Jochen Profit, Henry Lee, Sarah Kunz, Marinka Zitnik, Jure Leskovec, Ciaran Phibbs, Scott Lorch, Douglas Staiger, Jeannette Rogowski, and Jeffrey Gould.

\section*{Funding}
SVS and MS acknowledge funding support from the University of Vermont and Northeastern University. The funders played no role in the study design. JDH and LTG are employees of Vermont Oxford Network. EME is an employee of the University of Vermont funded by Vermont Oxford Network.

\section*{Authors' information}
Not applicable

\bibliographystyle{ieeetr}

\end{document}